\documentclass[traditabstract]{aa} 
\usepackage{graphicx}
\usepackage{txfonts}
\usepackage{natbib}
\usepackage{color}
\usepackage{multirow}
\usepackage{lscape}
\usepackage{latexsym}
\usepackage{changepage}         
\bibpunct{(}{)}{;}{a}{}{,} 

\usepackage[usenames,dvipsnames,svgnames,table]{xcolor}
\usepackage[breaklinks, colorlinks, citecolor=CornflowerBlue]{hyperref}

\begin{document}

\definecolor{Red}{rgb}{1,0,0}
\authorrunning{Remco F.J. van der Burg, et al.}
   \title{The abundance of ultra-diffuse galaxies from groups to clusters}
        \subtitle{UDGs are relatively more common in more massive haloes}
   \author{Remco~F.~J.~van der Burg\inst{1,2}, Henk Hoekstra\inst{3}, Adam Muzzin\inst{4}, Crist\'obal Sif\'on\inst{5}, Massimo Viola\inst{3},\\
        Malcolm~N.~Bremer\inst{6}, Sarah~Brough\inst{7}, Simon~P.~Driver\inst{8,9}, Thomas~Erben\inst{10}, Catherine~Heymans\inst{11}, Hendrik~Hildebrandt\inst{10}, Benne~W.~Holwerda\inst{12}, Dominik~Klaes\inst{10}, Konrad~Kuijken\inst{3}, Sean~McGee\inst{13}, \\Reiko~Nakajima\inst{10}, Nicola~Napolitano\inst{14}, Peder~Norberg\inst{15}, Edward~N.~Taylor\inst{16}, Edwin~Valentijn\inst{17}}

           \institute{IRFU, CEA, Universit\'e Paris-Saclay, F-91191 Gif-sur-Yvette, France
         \\  \email{remco.van-der-burg@cea.fr}  
                \and Universit\'e Paris Diderot, AIM, Sorbonne Paris Cit\'e, CEA, CNRS, F-91191 Gif-sur-Yvette, France
              \and Leiden Observatory, Leiden University, P.O. Box 9513, 2300 RA Leiden, The Netherlands
\and Department of Physics and Astronomy, York University, 4700 Keele St., Toronto, Ontario, Canada, MJ3 1P3
\and Department of Astrophysical Sciences, Peyton Hall, Princeton University, Princeton, NJ 08544, USA
        \and Astrophysics Group, School of Physics, University of Bristol, Tyndall Avenue, Bristol BS8 1TL, England
\and School of Physics, University of New South Wales, NSW 2052, Australia
        \and International Centre for Radio Astronomy Research (ICRAR), The University of Western Australia, 35 Stirling Highway, Crawley, WA6009, Australia
        \and School of Physics \& Astronomy, University of St Andrews, North Haugh, St Andrews, KY16 9SS, UK
        \and     Argelander-Institut f\"ur Astronomie, Auf dem H\"ugel 71, 53121 Bonn
   \and Scottish Universities Physics Alliance, Institute for Astronomy, University of Edinburgh, Royal Observatory, Blackford Hill, Edinburgh, EH9 3HJ, UK
        \and Department of Physics and Astronomy, University of Louisville, Louisville KY 40292 USA
        \and School of Physics and Astronomy, University of Birmingham, Edgbaston, Birmingham B15 2TT, England
        \and INAF-Osservatorio Astronomico di Capodimonte, via Moiariello 16, 80131 Napoli, Italy
                \and ICC \& CEA, Department of Physics, Durham University, South Road, Durham DH1 3LE, UK
        \and Centre for Astrophysics and Supercomputing, Swinburne University of Technology, Hawthorn 3122, Australia
                \and Kapteyn Institute, University of Groningen, PO Box 800, NL 9700 AV Groningen
        }

   \date{Submitted 8 June 2017 ; Accepted 7 August 2017}

  \abstract{In recent years, many studies have reported substantial populations of large galaxies with low surface brightness in local galaxy clusters. Various theories that aim to explain the presence of such ultra-diffuse galaxies (UDGs) have since been proposed. A key question that will help to distinguish between models is whether UDGs have counterparts in host haloes with
lower masses, and if so, what their abundance as a function of halo mass is. We here extend our previous study of UDGs in galaxy clusters to galaxy groups. We measure the abundance of UDGs in 325 spectroscopically selected groups from the Galaxy And Mass Assembly (GAMA) survey. We make use of the overlapping imaging from the ESO Kilo-Degree Survey (KiDS), from which we can identify galaxies with mean surface brightnesses within their effective radii down to $\sim$25.5 mag arcsec$^{-2}$ in the $r$ band. We are able to measure a significant overdensity of UDGs (with sizes $r_{\mathrm{eff}} \geq 1.5$ kpc) in galaxy groups down to $M_{200}=10^{12}\,\mathrm{M_{\odot}}$, a regime where approximately only one in ten groups contains a UDG that we can detect. We combine measurements of the abundance of UDGs in haloes that cover three orders of magnitude in halo mass, finding that their numbers scale quite steeply with halo mass: $\mathrm{N_{UDG}} (\mathrm{R}<R_{200}) \propto M_{200}^{1.11\pm0.07}$. To better interpret this, we also measure the mass-richness relation for brighter galaxies down to $M^{*}_{r}+2.5$ in the same GAMA groups, and find a much shallower relation of $\mathrm{N_{Bright}} (\mathrm{R}<R_{200}) \propto M_{200}^{0.78\pm0.05}$. This shows that compared to bright galaxies, UDGs are relatively more abundant in massive clusters than in groups. We discuss the implications, but it is still unclear whether this difference is related to a higher destruction rate of UDGs in groups or if massive haloes have a positive effect on UDG formation.}
   \maketitle
%

\hyphenation{in-tra-clus-ter}
\hyphenation{rank-or-der}

\section{Introduction}
While early studies of low surface brightness{} galaxies date back several decades \citep[e.g.][]{impey88,turner93,dalcanton97}, the discovery of a substantial number of very large diffuse galaxies in local clusters has drawn significant attention in recent years. \citet{vandokkum15} identified a significant population of large ($r_{\mathrm{eff}} \geq 1.5$ kpc) galaxies with low central surface
brightness ($\mu(g,0)=24-26\,\mathrm{mag\,arcsec^{-2}}$) in the Coma cluster, and dubbed them ultra-diffuse galaxies (UDGs). 

\bigskip
In the two years since, several studies reported populations of UDGs in different galaxy cluster samples. Some of the studies focused on the nearby and well-studied Fornax \citep{munoz15}, Coma \citep{koda15, kadowaki17} and Virgo \citep{mihos15, davies16} clusters. Other works studied more distant clusters \citep{vdB16, roman16a}, including a Frontier-fields cluster at $z=0.3$ \citep{janssens17}. It is clear that UDGs are a common phenomenon in galaxy clusters, and their ubiquity has to be understood in the context of existing galaxy formation models.

Several explanations have been proposed in the literature to explain this class of galaxies. \citet{vandokkum15} suggested that UDGs are some sort of ``failed'' galaxy, with a relatively massive dark matter halo like the one expected for an $L^{*}$ galaxy. Such a high mass would make them capable of surviving very close \citep[down to only $\sim$ 300 kpc, see][]{vdB16} to the cluster centres. They may have stopped forming stars due to feedback processes, which is again possibly related to a massively overdense environment. \citet{yozinbekki15} were indeed able to reproduce several observed properties of UDGs with a simple tidal disruption model that linked their presence exclusively to dense environments.

There are also less exotic explanations. \citet{amorisco16} claimed that UDGs can be readily explained by a standard model of disk formation, and that UDGs are simply the extremes of a continuous distribution in size and luminosity. This would mean that they also exist in lower-density environments, albeit with possibly more disk-like morphologies. \citet{rong17} reached a similar conclusion based on semi-analytic galaxy formation models. \citet{dicintio17} used the NIHAO simulation to study the formation of UDG-type galaxies, and they also suggested that internal processes, particularly outflow-driven feedback, may be responsible for their formation. Feedback from active galactic nuclei in dwarf galaxies may also contribute to their diversity and range in morphologies \citep{silk17}. 

Such opposing formation scenarios make different, observationally testable, predictions. One way forward is to investigate the failed-galaxy scenario by measuring the halo masses of UDGs. \citet{beasley16} estimated the mass of a UDG in the Virgo cluster based on the velocity dispersion of globular clusters that are associated with this galaxy, finding that they are overly massive for their stellar mass. Several other studies also based their mass estimates on the specific frequency, or number, of globular clusters in UDGs, reaching contradictory conclusions \citep{amorisco16b,penglim16,beasley16b}. \citet{dokkum16} measured a velocity dispersion of UDG DF44, which, upon extrapolation to $M_{200}$, is consistent with a halo mass of $M_{200}\sim 10^{12}\, \mathrm{M_{\odot}}$. In contrast, \citet{sifon17} measured the average mass associated with UDGs directly using weak gravitational lensing, finding that this is too low for them all to be ``failed'' massive galaxies. It is likely that the range of mass measurements of individual UDGs is real, and not (only) due to different assumptions in the analyses of different studies. This may hint at a range of different formation scenarios for UDGs \citep[e.g.][]{zaritsky17}. 

Another critical observational study to constrain the formation scenario of UDGs is to measure their abundance in different environments. \citet[][hereafter vdB16]{vdB16} studied eight clusters that span an order of magnitude in halo mass, and found that the number of UDGs per cluster scales as $N_{\mathrm{UDG}}\propto M_{200}^{0.93\pm0.16}$, where $M_{200}$ is the virial mass of the host halo. 
At even lower halo masses, \citet{roman16b} were able to identify UDGs in three nearby groups by making use of three-band photometric data from the SDSS deep Stripe 82. Their work shows that UDGs are also present in galaxy groups, with an abundance that is close to what is expected from extending the vdB16 relation down to lower masses. We note, however, that the sample they studied are the three Hickson Compact Groups \citep[][]{hickson82} that overlap with the SDSS Stripe 82, and this may not be fully representative of the underlying population of groups. 

In this work we perform a systematic search for UDGs in galaxy groups, considering the 197 deg$^2$ overlap region between the optical-imaging Kilo-Degree Survey (KiDS) with the groups identified with the Galaxy And Mass Assembly (GAMA) spectroscopic survey. The groups are selected completely independently of their possible UDG content. By performing an analysis that is close to the one presented in vdB16, this allows us to study the abundance of UDGs in a consistent way over a wide halo mass range. A previous search for galaxies with low surface brightness was performed using SDSS imaging of the same GAMA region \citep{williams16}. Based on the locations of these galaxies, it was shown that they are plausibly associated with the $z<0.1$ large-scale structure \citep[Fig.~6 in][]{williams16}. In this paper we quantify this statement, focusing on the GAMA groups, based on the standardised galaxy selection using the UDG criteria on the KiDS imaging.

The paper is organised as follows. Section~\ref{sec:dataoverview} provides an overview of the data products from the KiDS and GAMA surveys we used. Section~\ref{sec:analysis} describes the selection of our sample of UDGs, and how we associated them with the galaxy
groups in a statistical sense. Section~\ref{sec:results} presents our main results: the abundance of UDGs as a function of halo mass, and the size distribution of the UDG sample. We discuss our findings in Sect.~\ref{sec:discussion}, compare them to previous work and to a measurement of the mass-richness relation for the same GAMA groups to help to interpret our results. We summarise in Sect.~\ref{sec:conclusion}. Appendix \ref{sec:robustness} lists several robustness tests that we performed to increase our confidence in the presented results. 

All magnitudes we quote are in the AB magnitudes system, and we adopt angular-diameter and luminosity distances corresponding to $\Lambda$CDM cosmology with $\Omega_{\mathrm{m}}=0.3$, $\Omega_{\Lambda}=0.7$ and $\mathrm{H_0=70\, km\, s^{-1}\,  Mpc^{-1}}$. Unless stated otherwise, error bars denote 1$\sigma$ uncertainties, or 68\% confidence intervals. 

\section{Data}\label{sec:dataoverview}
Our UDG candidates are identified using photometric data from the Kilo-Degree Survey \citep[KiDS,][]{kidsdejong15}, and the environments (in particular group properties) are characterised using the Galaxy And Mass Assembly \citep[GAMA,][]{drivergama11} group catalogue. In this study, we focus on the overlap region between the KiDS and GAMA surveys, totalling an effective (unmasked) area of 170 square degrees, and the surveys are summarised in turn below. 

\begin{figure*}
\begin{center}
\resizebox{0.9\hsize}{!}{\includegraphics{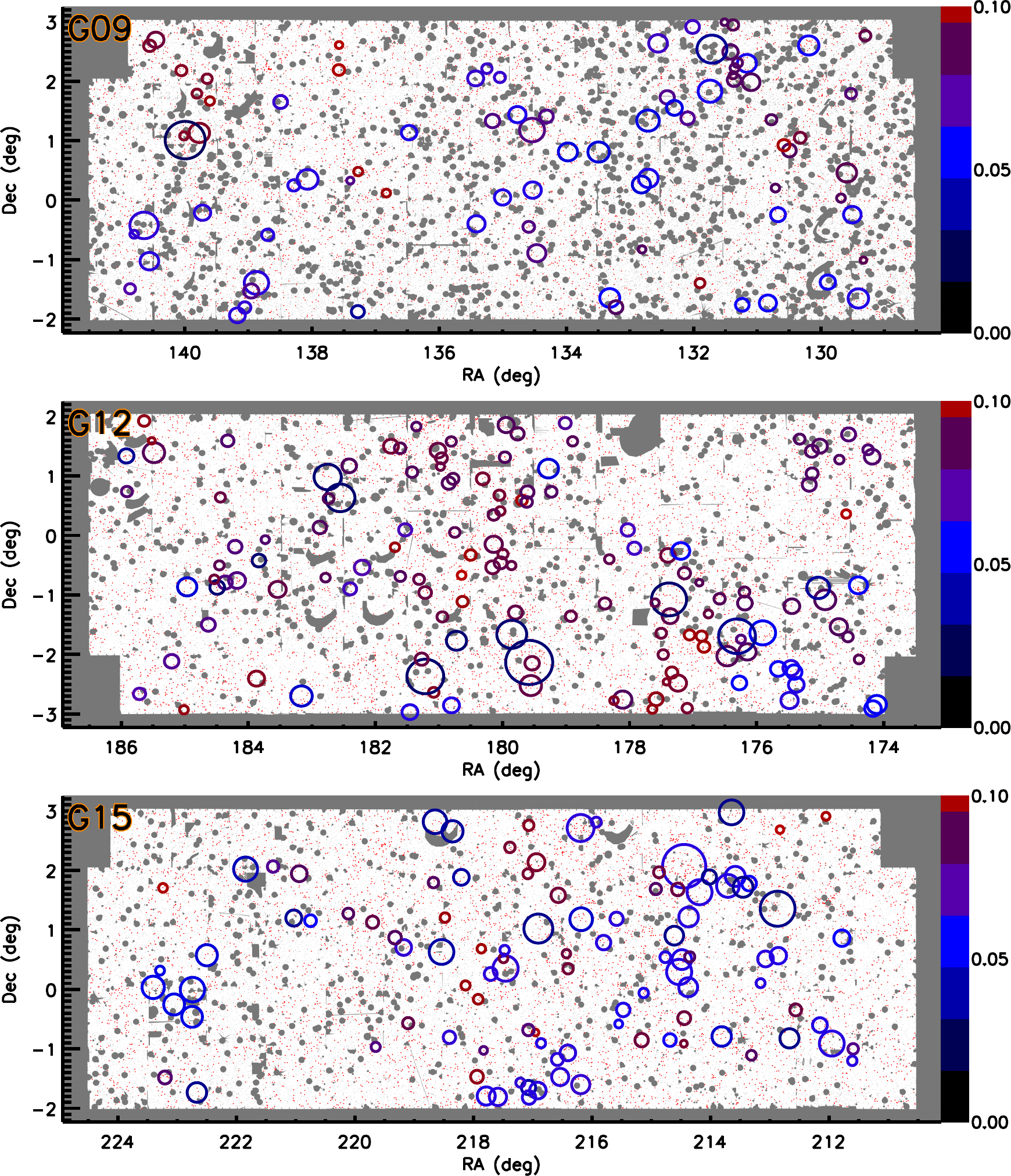}}
\caption{Three KiDS+GAMA fields. The combined KiDS area is shown in white, with regions masking bright stars, reflective haloes, and image artefacts shown in grey. Circles indicate the actual angular sizes of $R_{200}$ disks of GAMA groups up to $z\leq 0.10$, with colours referring to their redshifts (cf.\ colour scale on the right). Small red dots show the positions of all UDG candidates that satisfy our final selection criteria.}
\label{fig:gama_allfields_overview}
\end{center}
\end{figure*}

\begin{table}
\caption{Rough overview of the three KiDS+GAMA fields studied here. Each KiDS tile covers roughly 1 deg$^2$. The effective areas are reduced by $\sim 15\%$ due to the masking, see Fig.~\ref{fig:gama_allfields_overview} for a graphical representation.}
\label{tab:dataoverview}
\begin{center}
\begin{tabular}{l r r r r}
\hline
\hline
Field & KiDS tiles & Eff area [deg$^2$] & UDG candidates & Groups \\
\hline
G09&          64&50.7&        4635&          83\\
G12&          64&56.2&        5815&         140\\
G15&          69&62.5&        6145&         102\\
\hline
Total&         197&169.5&       16595&         325\\
\hline
\end{tabular}
\end{center}
\end{table}

\subsection{Kilo-Degree Survey}
 KiDS is an on-going ESO wide-area imaging survey in the $ugri$
bands that will cover 1500 deg$^2$, and which is designed to map the matter distribution in the Universe using weak gravitational lensing. Here we used the `KiDS-450' data release, which covers approximately 450 deg$^2$ \citep{hildebrandt17, dejong17}.

Our starting point is the KiDS $r$-band imaging, reduced and stacked with the THELI \citep{erben13} pipeline. The data are taken under dark conditions with minimum atmospheric extinction. To facilitate weak-lensing measurements in the $r$ band, they are taken in conditions with excellent seeing (PSF FWHM$<0.8''$). Stacks are composed of five dithers with a combined integration time of 1800s. 

The stacks are designed for optimal weak-lensing shape measurements, and there are still residual background patterns that we have to account for in this study. We subtracted these residual background effects using a mesh size of 150 pixels ($\sim32''$), smoothed with a median filter of size 5$\times$5 meshes. This corrects background structures on a scale that is much larger than the sizes of the galaxies we study. 

We used the conservative masks designed by the KiDS team, which mask $\sim15\%$ of the total survey area. This includes bright stars, their reflective haloes, and image artefacts \citep[for more details see][]{kuijken15}. Figure~\ref{fig:gama_allfields_overview} shows the unmasked area of the three KiDS fields we consider in this study in white. 

\subsection{Galaxy And Mass Assembly (GAMA)}
GAMA \citep{drivergama11,liske15} is a spectroscopic survey that is complete down to an $r$-band magnitude limit of 19.8. Each GAMA field is visited several times to overcome sampling problems that are due to close galaxy pairs, which are common in groups and clusters of galaxies. Using a friends-of-friends (FoF) algorithm on the spectroscopic catalogue, \citet{robotham11} have constructed a group catalogue for the GAMA survey regions that links $\sim 14\%$ of the GAMA galaxies to a group with at least five spectroscopic members. Since this first version of the catalogue (G$^3$Cv1), the linking scheme has been updated several times. Including the current version of the spectroscopic catalogue, G$^3$Cv9 is now the most recent version of the group catalogue. While it covers a total of five survey fields, we focus on the three Equatorial GAMA fields G09, G12, and G15, which overlap with KiDS (cf.~Table~\ref{tab:dataoverview}). We considered all groups from G$^3$Cv9 in these survey fields that have at least five FoF members and are in the redshift range $0.01 \leq z\leq 0.10$. This selection resulted in a total of 325 groups, see Table~\ref{tab:dataoverview}. 

Several group mass definitions are in use within the GAMA collaboration. Originally, group masses were measured dynamically, using the spectroscopic group members as tracers of the potential. While this dynamical method is expected to work reasonably well for groups with a large number of members, these masses are unreliable in the low-mass regime, however, where there are few members \citep{robotham11}. Instead, the masses used in this work are based on the total group luminosity in the $r$ band.
\citet{viola15} have performed a weak-lensing study of the overlap region between KiDS and GAMA to measure the total mass associated with groups, in bins of their total $r$-band luminosity. We used the almost-linear relation they found between total luminosity and weak-lensing mass to estimate $M_{200}$ for each group. The $R_{200}$ is defined, as usual, as the radius corresponding to a sphere that contains this mass. The group centres are defined as their $r$-band luminosity-weighted centre of mass. 

Figure \ref{fig:gama_allfields_overview} gives an overview of the three GAMA regions studied here. The white area is the (unmasked) KiDS area considered here. The circles mark the locations of groups, with colours indicating their redshifts, and their sizes are the projected angular size of their $R_{200}$ disks. 
An additional criterion to define our group sample is that less than 50\% of their $R_{200}$ disk is masked in the KiDS images (which removes 17 groups from the group sample, bringing the total to 325). 

\section{Analysis}\label{sec:analysis}
Ultra-diffuse galaxies are defined in physical units \citep[$r_\mathrm{{eff}}\geq$ 1.5 kpc,][]{vandokkum15} and have a maximum size of about $r_\mathrm{{eff}}\sim7.0$ kpc (vdB16). Since we here consider GAMA groups over a wide redshift range, these physical sizes correspond to a range of different angular sizes. For example, at $z=0.01$, we would have to consider galaxies with angular sizes between $7\farcs3$ and 34$''$, while at $z=0.10,$ this would correspond to sizes between $0\farcs8$ and $3\farcs8$. 
From an observational point of view, we have to work in angular units, which initially leads to a redshift-dependent selection. However, when we exploit the wide range of redshifts of the GAMA groups, we can make definitive statements about their UDG content, as outlined below.

\subsection{Large galaxies with low surface brightness in KiDS}\label{sec:selectioncriteria}
Although we work in angular units, we performed an analysis that is as close as possible to the analysis presented in vdB16, where data from the Canada-France-Hawaii Telescope (CFHT) were analysed. 
As a first step, we used \texttt{SExtractor} \citep{bertinarnouts96} on the $r$-band stack to detect sources, using a Gaussian filter with a FWHM of 5 pixels ($\sim 1''$ with our detector) to improve our sensitivity towards faint extended galaxies. The KiDS pixel scale is slightly coarser than CFHT MegaCam (0.214$''$/pix versus to 0.185$''$/pix), so we require sources to have at least 15 adjacent pixels that are at least 0.90 sigma above the background. These choices keep the purity of detecting real objects (and not noise fluctuations) high (cf.~vdB16). We detected a total of $12\,243\,224$ sources in the KiDS fields. 

We applied liberal selection criteria to these \texttt{SExtractor} output catalogues to perform a first selection of large extended galaxies (and thus potential UDGs at the redshifts of the GAMA groups). We selected against stars and other compact objects, requiring that $r_{2''} > 0.9 + r_{7''}$, where $r_{x''}$ is the $r$-band magnitude within a circular diameter of $x$ arcsec. This pre-selection was merely intended to speed up the overall analysis. We performed several tests to ensure that this cut did not affect the final selection of galaxies that we aimed to study, even if a fraction of UDGs is nucleated \citep[cf.][]{yagi16}. After applying the KiDS survey mask, we were left with $873\,702$ sources in the considered KiDS area. 

These sources were our input to \texttt{GALFIT} \citep{peng02}, which we used to estimate morphological parameters while masking any pixels belonging to neighbouring sources. We then selected galaxies with best-fit S\'ersic parameters in the range $24.0 \leq \langle\mu(r,r_\mathrm{eff})\rangle \leq 26.0 \, \mathrm{mag\, arcsec^{-2}}$, circularised effective radii $1\farcs5\leq r_\mathrm{eff}\leq8\farcs0$ that are fit within 7 pixels from the \texttt{SExtractor} position, and have a S\'ersic index $n\leq$4. We note that the surface brightness criterion is somewhat shallower (26.0 versus 26.5) than the criterion applied in vdB16 because we work with shallower
data here. Some examples of selected galaxies are shown in Appendix~\ref{sec:examples}.

We find that \texttt{GALFIT} provides reasonable fits to the data in most cases; we therefore trust the \texttt{GALFIT} parameters, and the selection can be considered as being quite pure. We did not make any additional subjective or by-eye checks for individual sources, so that this study remains reproducible. We assume that any residual impurity does not correlate with the locations of groups and is eventually removed (in a statistical sense) by our background subtraction. Given that KiDS is untargeted, this is a reasonable assumption. To test the performance of our procedure and to gauge the completeness of our final UDG candidate selection, we also performed all processing steps on a set of tailored image simulations, as described next. 

\subsection{Image simulations}\label{sec:simulations}
\begin{figure}
\resizebox{\hsize}{!}{\includegraphics{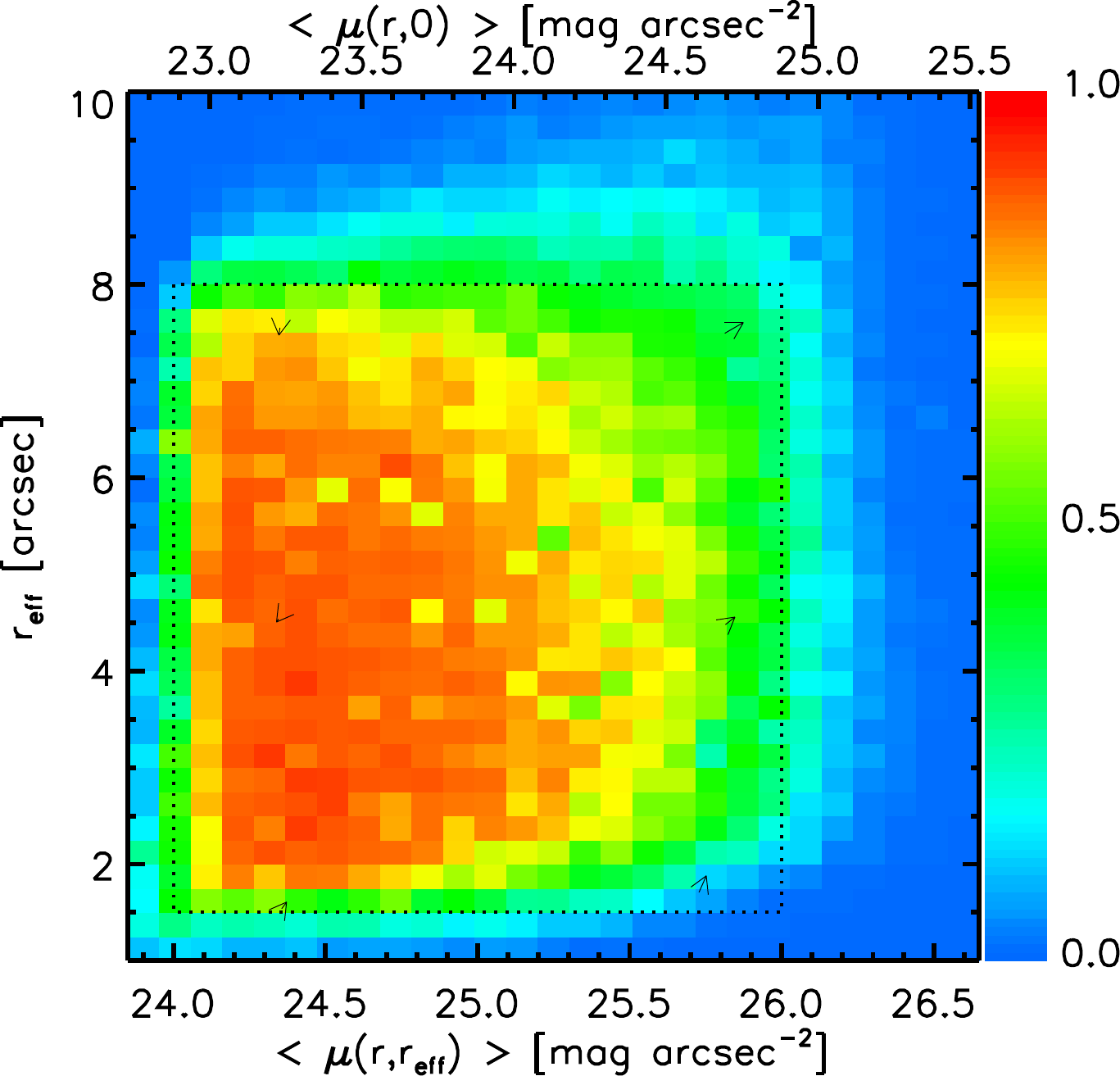}}
\caption{Recovered fraction of sources that were injected in the KiDS $r$-band stacks as a function of circularised effective radius, and the mean surface brightness within the effective radius. The dotted region defines the selection criteria used in this study. Arrows show the negligible bias in the morphological parameters recovered with \texttt{GALFIT} in six corners of parameter space.}
\label{fig:completeness}
\end{figure}

Following vdB16, we injected objects with S\'ersic profiles with $n$=1 \citep[typical values for UDGs, cf.][]{koda15,vdB16} at random positions in the KiDS $r$-band image stacks. Effective radii ($r_\mathrm{eff}$) were drawn uniformly between 1$''$ and 10$''$ (we worked in angular units here). Central surface brightnesses were drawn uniformly from the range $22.7 < \mu(r,0)/[\mathrm{mag\, arcsec^{-2}}] < 26.0$, and the ellipticity parameter $\epsilon=(1-q)/(1+q)$, where $q$ is the ratio of minor/major axis, was drawn uniformly between 0 and 0.2. Injecting 1500 sources per realisation, with 10 realisations per KiDS tile, we simulated several million sources. 

In the following we work in units of the mean surface brightness within the effective radius (following vdB16). This parameter is exactly 1.12 mag arcsec$^{-2}$ larger than the \textit{central} surface brightness for S\'ersic-profiles with $n$=1, and is more closely related to the detectability of a source. In particular, as noted in vdB16, the recovery rate of simulated sources with slightly different S\'ersic-indices ($n$=0.5 or $n$=1.5) is similar when expressed in these units. It is also noted there that the detectability of a source of a given effective (circularized) radius does not sensitively depend on its ellipticity. 

Figure~\ref{fig:completeness} shows the completeness of the simulated sources. The completeness is defined as the fraction of injected sources, in bins of size and surface brightness, that are 1) detected with \texttt{SExtractor}, 2) pass the pre-selection based on the \texttt{SExtractor} output, and 3) have best-fit morphological parameters corresponding to the selection criteria we outlined above. These simulated sources were thus processed through exactly the same pipeline as the ``real'' sources. While only 2\% of ``real'' sources that were fed to \texttt{GALFIT} satisfy these criteria, the recovery rate of simulated sources is $\sim$66\%. Most ``real'' sources that did not match the final selection criteria are either too small (the pre-selection criteria on the \texttt{SExtractor} output catalogues were very inclusive, as outlined above) or too bright (there was no pre-selection on brightness).

With our final selection criteria, we are left with 16\,595 large and faint galaxies in an unmasked KiDS area of 169.5 deg$^2$. We ensured that we did not count the same galaxies several times because of the slightly overlapping KiDS tiles. 
As a sanity check, we applied exactly the same selection criteria to the sources detected in the CFHTLS Deep fields (with noise added to resemble the cluster imaging used in vdB16). We found 343 sources that satisfy these criteria (in angular units) in a combined effective area of 3.52 deg$^2$. This means that the source densities are remarkably similar between the two surveys, providing further confidence that we can compare results from these different instrumental setups and depths.

\subsection{Associating UDGs with GAMA groups}\label{sec:analysis_correctionfactors}
We now link the selected sample of galaxies to UDGs at the redshifts of the GAMA groups. As in vdB16, we considered UDGs with physical sizes between 1.5 kpc and 7.0 kpc. We note that for galaxy groups with redshift $z>0.051$, we did not probe the smallest UDGs (because they would have an angular size smaller than $1\farcs5$). Correspondingly, for galaxy groups with redshift $z<0.044$, we did not probe the largest UDGs (because they have an angular size larger than $8\farcs0$). As described below, we combined information from all GAMA groups, exploiting the entire redshift range of the GAMA groups ($0.01\leq z\leq 0.10$), to turn our samples of galaxies with sizes between $1\farcs5$ and $8\farcs0$ into physical sizes (and thus UDGs, after accounting for fore- and background interlopers). Given that only 10\% of the current age of the Universe has passed since $z=0.10$, we assume that the UDG content of the galaxy groups does not evolve over this redshift range. We proceeded as follows:
 
\begin{itemize}
\item For each GAMA group, we counted the number of large galaxies
with low surface brightness that satisfy our \textit{angular} selection criteria (cf.~Fig.~\ref{fig:completeness}), the \textit{physical} size criteria corresponding to the group redshift, and that are projected within $R_{200}$ from the group centre. We estimated the background value on this number, namely the counts expected from the average number density of the particular KiDS field studied here (i.e.~G09, G12, or G15)\footnote{In principle, to properly estimate the background, structures that are physically associated with the group in question should be masked out before estimating the background. However, given the very low overdensities of UDGs in these groups together with the large statistical uncertainties in this work, we find that even in an extreme case, namely when we mask all \textup{}groups up to $z\leq 0.10$, this has a negligible effect, namely at most 10\% of the statistical uncertainty, on the final results (cf.~Appendix~\ref{sec:robustness}). The same is true when we estimate the background for groups using the average over all the fields. In the following we therefore use a simple average number density per field to estimate the background.}. The raw counts minus the background expectation and associated Poisson errors are our main measurements. The total overdensity of measured UDGs in all GAMA groups is 29\% compared to the background. 

\item We corrected this number for the masked fraction of the $R_{200}$ disk by assuming that the UDG density is homogeneous. The average correction factor for all groups is 17\% to the number counts, which is close to the masked fraction of the entire KiDS survey area. We recall that we did not consider groups in which more than half the $R_{200}$ disk is masked.

\item We corrected this number for incompleteness, using our extensive set of image simulations (as described in Sect.~\ref{sec:simulations}). We counted the number of simulated galaxies with \textit{intrinsic} parameters in our selection region, and divided this by the number of simulated galaxies with \textit{measured} parameters in our selection region. Key here is that a larger region of parameter space is simulated than we considered for the analysis. This means that scatter between intrinsic and measured parameters can be taken into account. For this step we had to assume an underlying intrinsic distribution in morphological parameters. We started with the values measured for the cluster population of UDGs in vdB16; a steep size distribution given by $n \mathrm{[dex^{-1}]}\propto r_\mathrm{eff}^{-3.4}$, and a homogeneous distribution in surface brightness (in this region of parameter space). This led to correction factors ranging from 0.57 at the lowest redshift to 1.50 at $z=0.10$. 

\item Since we only considered sources with sizes between $1\farcs5$ and $8\farcs0$ arcsec, we estimated which fraction of UDGs with physical sizes between 1.5 and 7.0 kpc we missed in this count. For this we again used the size distribution. There is no correction (correction factor=1) in the range $0.044<z<0.051$, since all physical sizes are represented in that redshift range. It increases only slightly towards lower redshift (since large UDGs are rare), but peaks at 8.3 for groups at $z=0.10$. This correction factor depends on the assumed UDG size distribution, which we iterated over in Sect.~\ref{sec:sizedist}. In Appendix~\ref{sec:robustnesssize} we show that our final results are not significantly affected by this assumed size distribution. 

\item Since vdB16 were able to consider a slightly broader range in mean surface brightness within the effective radius ($24.0 \leq \langle\mu(r,r_\mathrm{eff})\rangle \leq 26.5 \, \mathrm{mag\, arcsec^{-2}}$, instead of up to 26.0), we increase our counts by 25\%, assuming that the distribution is homogeneous in surface brightness (vdB16, and supported by the current analysis, see Fig.~\ref{fig:sizedist}). We note that this correction was
made only for the purpose of comparing our results to those from vdB16 and other literature studies. 
\end{itemize}

\section{Results}\label{sec:results}
The next two subsections present our main results, namely the abundance of UDGs as a function of halo mass, and their size distribution in the GAMA groups. They are measured iteratively, since the result of each of the two measurements potentially influences the other. We explain the reason for this interdependence in Appendix~\ref{sec:robustness}, and conclude there that each measurement is robust and only mildly influenced by the other (i.e.~to a degree that is smaller than the size of the statistical uncertainties). 

\subsection{Abundance of UDGs in galaxy groups}\label{sec:abundancehalomass}

\begin{table}
\caption{Average number of UDGs per group in bins of halo mass. Bins have a width of 0.4dex in halo mass, starting at $10^{12}\,\mathrm{M_{\odot}}$. These are the data points shown in Fig.~\ref{fig:abundance1}. Two types of uncertainty are given: first the Poisson uncertainty on the average value in each bin, second the bootstrap uncertainty from resampling groups in each bin (drawing with replacement). The latter uncertainty also includes the Poisson uncertainty on individual groups, and is therefore never smaller than the former (and they are equal when there is only one group in a bin). The last column lists as a comparison the richness of bright galaxies in the same groups and bins.}
\label{tab:abundance}
\begin{center}
\setlength{\extrarowheight}{0.05cm}
\begin{tabular}{c r r r}
\hline
\hline
Binmean & Ngroups & UDGs per group & Richness\\ 
log$_{10}[M_{200}/\mathrm{M_{\odot}}]$&&&$M_{r} < M_{r}^{*} +2.5$\\
\hline
12.23&8&$-0.21^{+0.51+0.52}_{-0.45-0.61}$&$1.80^{+0.50+0.40}_{-0.44-0.78}$\\
12.62&38&$0.32^{+0.62+1.08}_{-0.55-0.93}$&$2.93^{+0.30+0.39}_{-0.27-0.40}$\\
13.03&139&$1.38^{+0.56+1.26}_{-0.49-1.04}$&$4.69^{+0.20+0.30}_{-0.18-0.26}$\\
13.36&107&$3.31^{+0.89+1.51}_{-0.78-1.62}$&$7.98^{+0.31+0.45}_{-0.27-0.48}$\\
13.77&27&$7.54^{+2.52+5.22}_{-2.21-3.97}$&$19.31^{+0.94+1.97}_{-0.83-1.59}$\\
14.04&3&$44.93^{+13.95+21.54}_{-12.23-38.98}$&$39.60^{+3.93+5.22}_{-3.44-7.79}$\\
14.52&1&$64.29^{+18.16+18.16}_{-15.92-15.92}$&$81.84^{+9.66+9.66}_{-8.47-8.47}$\\
\hline
\end{tabular}
\end{center}
\end{table}

\begin{figure}
\resizebox{\hsize}{!}{\includegraphics{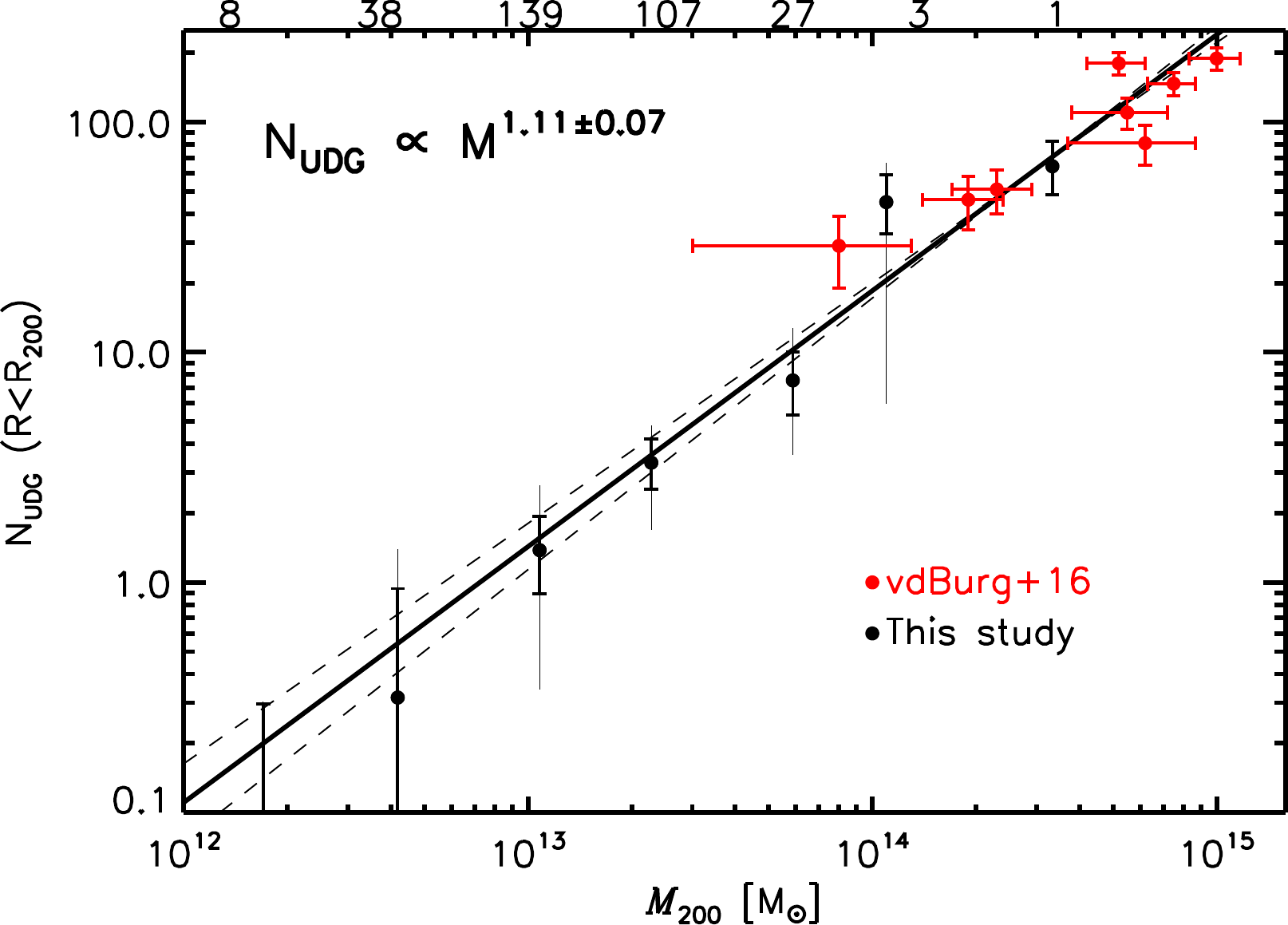}}
\caption{Abundance of UDGs as a function of halo mass. \textit{Thick black points:} averages in bins of halo mass. Two types of error are shown: thick errors with hats are Poisson uncertainties, while thin errors show bootstrap errors after resampling the groups in each bin. \textit{Thick red points:} values measured for eight individual clusters in vdB16. \textit{Black lines:} best-fitting power-law relation including all data points shown here. \textit{Dashes:} relations corresponding to the $\pm 1$-$\sigma$ uncertainties on the power-law index. The numbers on the top axis indicate the number of groups in each bin.}
\label{fig:abundance1}
\end{figure}

Figure~\ref{fig:abundance1} shows the abundance of UDGs as a function of halo mass for the GAMA groups. Group masses are based on a weak-lensing calibration in bins of their total $r$-band luminosity \citep{viola15}. Two types of error bars are shown, namely Poisson uncertainties and bootstrap errors where groups in each bin have been resampled (drawing with replacement). Data points are also presented in Table~\ref{tab:abundance}. We note that the statistical background subtraction may in principle yield negative values for the abundance of UDGs in groups; in this case, for the bin with few low-mass systems that have a low overdensity compared to the background. In Fig.~\ref{fig:abundance1} we also show the measurements from vdB16, which were obtained from a similar analysis, but are based on individual clusters. We note that halo masses are determined differently between vdB16 and the current study. vdB16 measured cluster masses dynamically \citep[cf.][]{sifon15}, while our group masses are based on their total $r$-band luminosity. However, the group masses are calibrated directly using weak gravitational lensing \citep{viola15}, while the dynamical masses of the clusters studied in vdB16 are tightly correlated with lensing mass (Herbonnet et al., in prep.). In Appendix~\ref{sec:robustness} we perform a conservative test to ensure that residual uncertainties in mass calibration between the two studies do not affect our results.

We fitted\footnote{We used a $\chi^2$ fitting recipe on data with errors in both coordinates, as described in Sect.~15.3 of \citet{numericalrecipesinc}.} a power-law relation between the halo mass and UDG abundance, combining both the cluster measurements from vdB16, and the binned group abundances presented here. For the clusters we took errors on individual halo masses into account, while for the KiDS+GAMA points we used the Poisson uncertainties, and we assumed an uncertainty of 0.1 dex in mean halo mass for each bin while performing the fit. A power law of the form 
\begin{equation}\label{eq:udgs}
N_{\mathrm{UDG}}=(19\pm2) \cdot \left[ \frac{M_{200}}{10^{14}\,\mathrm{M_{\odot}}}\right]^{1.11\pm0.07}
\end{equation}
provides a good description of the data points, with $\chi ^2/\mathrm{d.o.f.}=16.0/13$. 

We note that as is customary when measuring the number of galaxies in a halo (e.g.~richness), the numbers we quote are measured in a cylinder with radius $R_{200}$. For the number of galaxies within a sperical volume with radius $r_{200}$, there is a deprojection factor to be considered. Assuming that galaxies to first order \citep[but see][]{vdB15} trace a Navarro-Frenk-White (NFW) profile \citep{NFW}, the deprojection factor depends on the concentration of the halo. The typical concentration is a function of halo mass \citep[e.g.][]{duffy08,dutton14}, so that the deprojection factor ranges from about $\sim$0.80 for groups (assuming $c_{200}$=6), to $\sim$0.75 for clusters (assuming $c_{200}$=3). If we were
to deproject all values and repeat the fit, it would only change the best-fit slope from 1.11 to 1.10. Since we are primarily interested in the relative abundance as a function of halo mass and not in the absolute normalisation, we did not apply a deprojection in this work. 

\subsection{Size distribution of UDGs}\label{sec:sizedist}
\begin{figure}
\resizebox{\hsize}{!}{\includegraphics{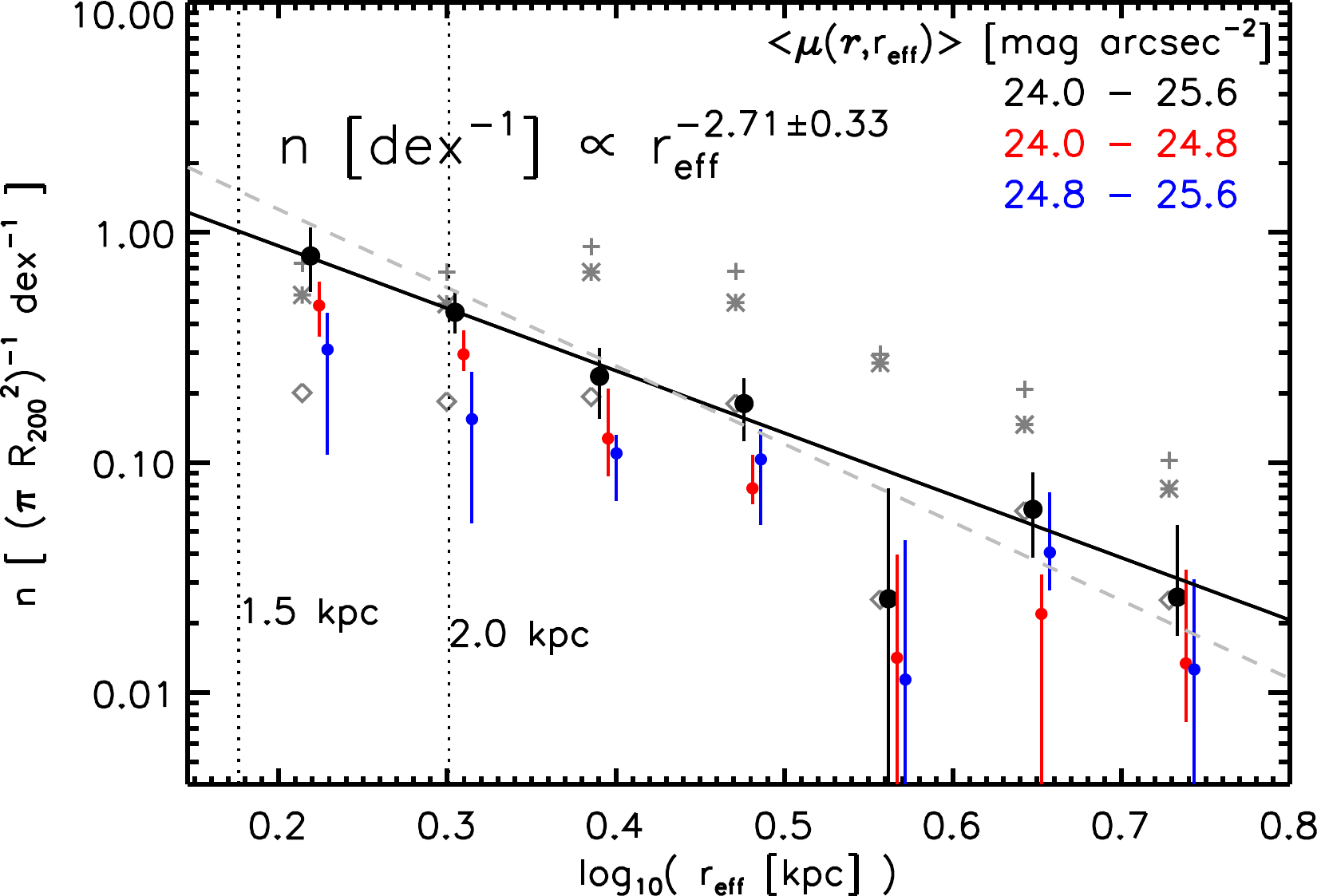}}
\caption{Measured size distribution of galaxies in the central projected $R_{200}$ for the GAMA groups. \textit{Grey pluses:} average raw counts per group. \textit{Grey note:} average values expected from fore- and background interlopers. \textit{Grey $ diamond$:} background-subtracted values. \textit{Black data points:} background-subtracted values, corrected for incompleteness, including bootstrap resampling errors. \textit{Blue and red points:} measured sizes for galaxies split over two bins of mean effective surface brightness, also after correction. In this size range, the data are well described by a power-law index of $-2.71\pm0.33$ (in bins of equal logarithmic size). \textit{Grey dashed line:} size distribution with power-law index of $-3.40$, which we found for UDGs in clusters (vdB16), arbitrarily normalised.}
\label{fig:sizedist}
\end{figure}

A related measurement to the abundance as a function of halo mass is the size distribution of UDGs in groups. As mentioned before (and explained in more detail in Appendix~\ref{sec:robustness}), both measurements are slightly interdependent and have to be performed iteratively. Figure \ref{fig:sizedist} shows the size distribution of UDGs that passed our main selection criteria, but have a stricter selection in surface brightness of $24.0 \leq \langle\mu(r,r_\mathrm{eff})\rangle \leq 25.6 \, \mathrm{mag\, arcsec^{-2}}$, since we are highly complete in this region of parameter space (cf.~Fig.~\ref{fig:completeness}). Each of the 325 groups contributes to the estimated size distribution, but the physical range that is covered depends on the redshift of
each group. Since more massive groups are generally found at higher redshift (where the volume probed is larger), we accounted for the fact that they contain more UDGs within their $R_{200}$ disk than lower-mass groups (cf.~Fig.~\ref{fig:abundance1}). We therefore scaled numbers by $M_{200}^{1.1}$, that is, by the relation found in Sect.~\ref{sec:abundancehalomass}. 

The drop in the background-subtracted counts (diamonds) at small physical sizes arises because groups at higher redshift do not contribute to these bins, and given that these are massive and therefore contain many UDGs, there is a large correction of up to a factor $\sim$5 for the smallest-size bin. After correcting the numbers and fitting a power-law distribution, we obtained a best fit $n \mathrm{[dex^{-1}]}\propto r_\mathrm{eff}^{-2.71\pm0.33}$ with $\chi^2/\mathrm{d.o.f.}=0.56$. We performed the same analysis on two subsamples with mean effective surface brightnesses of $24.0 \leq \langle\mu(r,r_\mathrm{eff})\rangle \leq 24.8 \, \mathrm{mag\, arcsec^{-2}}$ (shown in red) and $24.8 \leq \langle\mu(r,r_\mathrm{eff})\rangle \leq 25.6 \, \mathrm{mag\, arcsec^{-2}}$ (shown in blue), finding power-law indices that are identical within errors. The normalisation of each subsample is roughly 50\% of the full sample, indicating that the UDG distribution is roughly uniform as a function of mean effective surface brightness in this part of parameter space (similar to what was found by vdB16). 

\section{Discussion}\label{sec:discussion}
The measured size distribution of UDGs places powerful constraints on models that aim to explain the formation of UDGs \citep[e.g.][]{amorisco16}. The size distribution we find ($n \mathrm{[dex^{-1}]}\propto r_\mathrm{eff}^{-2.71\pm0.33}$) is very steep, indicating that the largest UDGs are rare. Thanks to the strong overdensity of the clusters studied in vdB16, the authors were able to measure the size distribution more precisely to be $-3.40\pm 0.19$ (see their Fig.~7). While the UDG size distribution is certainly steep, there is evidence (at the $2\sigma$ level) that the distribution depends on environment. It would be worthwhile to measure the size distribution specifically only in the lowest-mass groups, but we find that this results in such a low overdensity of UDG candidates with respect to the field that we cannot place tighter constraints on a possible environment dependence of the UDG size distribution in this work.

\subsection{Abundance of UDGs as a function of halo mass}\label{sec:discussionabundance}
The measured abundance of UDGs as a function of halo mass is perhaps even more important in constraining formation scenarios, as it directly addresses the question of \textit{\textup{where}} UDGs may have formed. We have measured the abundance now in a consistent manner over three orders of magnitude in halo mass. For the lowest-mass groups that enter into the analysis, we note that they contain on average fewer than one UDG per group. A power law of the form $N \propto M_{200}^{1.11 \pm 0.07}$ provides a good description of the abundance of UDGs as a function of their host halo mass. In Appendix~\ref{sec:robustness} we show that this measurement is robust with respect to the main assumptions that enter into the analysis. We note that the measurement presented here is more precise than the measurement performed by vdB16, where our measurement on a cluster sample spans only one order of magnitude in halo mass. vdB16 found a power-law exponent of $0.93\pm0.16$, which is consistent within $\sim 1 \sigma$ with this measurement. 

According to our best-fit relation, Eq.~\ref{eq:udgs}, roughly one in five systems like the Local Group, which has a combined mass of $M_{200}\approx 2\times 10^{12}\,\mathrm{M_{\odot}}$ \citep{gonzalezkravtsov14}, is expected to host a UDG within our selection boundaries. This rises to a substantial population of $\sim 200$ UDGs at halo mass scales of $M_{200}=10^{15}\,\mathrm{M_{\odot}}$. 

\begin{figure}
\resizebox{\hsize}{!}{\includegraphics{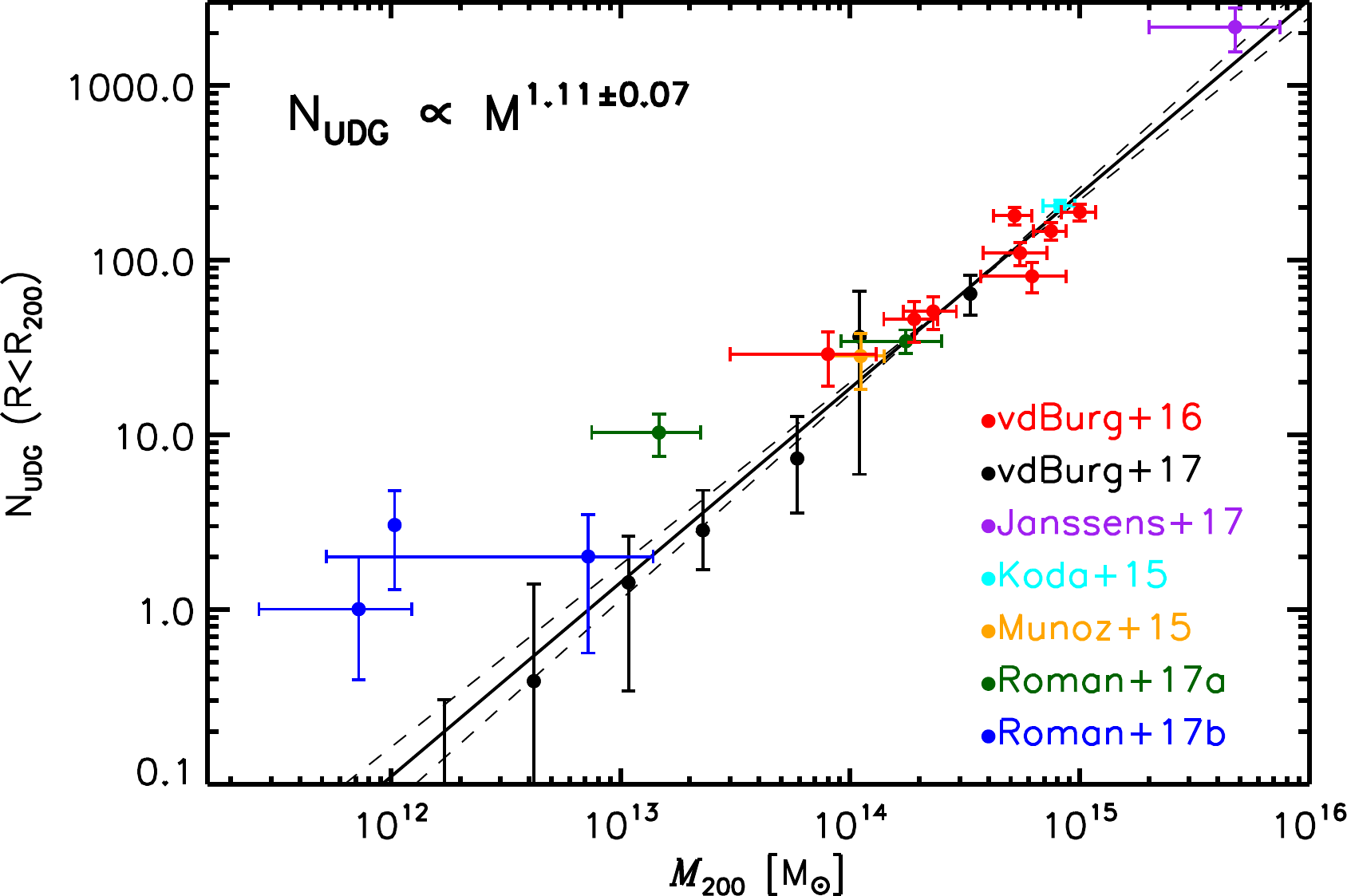}}
\caption{Same as Fig.~\ref{fig:abundance1}, but now expanding the plotted range, and showing literature measurements (as described in the text). Uncertainties shown here are the bootstrap error bars, from resampling groups within each bin.}
\label{fig:abundance2}
\end{figure}

We note that \citet{roman16b} measured the abundance of UDGs in three groups and found UDGs to be more abundant per unit halo mass in groups than in higher-mass systems (see Fig.~\ref{fig:abundance2} for a comparison), a conclusion that is different from ours. Low-number statistics may largely explain this apparent discrepancy. We note that many of the (especially low-mass) groups we studied in this work do not contain any UDGs, thus lowering the average to fewer than 1 UDG per group in the low-mass regime. Including Poisson scatter, the groups studied in \citet{roman16b} are fully consistent with being drawn from a sample comparable to ours.

There is another possible interpretation, related to the group properties themselves. The groups studied by \citet{roman16b} are the three Hickson Compact Groups \citep[HCGs,][]{hickson82},
which overlap with the SDSS Stripe 82. The HCG selection does not include loose groups, and is thus not representative for the diverse population of groups. In particular, constituent galaxies of HCGs have likely experienced only few merging events in their history \citep{proctor04,mendes05}. This may also have positively affected the survivability of UDGs in such groups. 

In Fig.~\ref{fig:abundance2} we also show the result from \citet{janssens17}, who measured the abundance of UDGs in a very massive cluster from the Hubble Frontier Fields. In Fig.~4 of \citet{janssens17} and in the accompanying text \citep[see also Fig.~6 in][]{roman16b}, a comparison is made with other measurements of the UDG abundance in different systems that have recently been presented in the literature. We include them here for reference, noting that most of them are consistent within 1$\sigma$ with the extrapolation of our best-fit relation, even at higher masses. Several UDGs have also been found to be possibly associated with a local galaxy group around NGC 5473 \citep{merritt16}. However, the surface brightnesses of these UDGs are too faint to be detectable with the depth of the KiDS imaging. We therefore did not include these results in Fig.~\ref{fig:abundance2}. The difference in surface brightness limits is a general caveat when comparing results from different studies. For instance, \citet{mihos15} studied three UDGs with surface brightnesses that are $\sim$2 magnitudes fainter than we can detect in the KiDS imaging. Figure~\ref{fig:abundance2} \citep{roman16b,janssens17} only includes studies with detection limits comparable to ours.

The power law we find here suggests that the number of UDGs per unit of host halo mass is increasing towards higher-mass haloes. It is perhaps even more insightful to compare the abundance of UDGs to the abundance of more massive galaxies in group- and cluster-sized haloes. Measuring the so-called mass-richness relation has been the subject of multiple studies in the past decades \citep{marinoni02,lin04,rines04,muzzin07b,andreon10b}. With very few exceptions \citep{kochanek03,uitert16}, they all consistently measured a relation $N\propto M^{\alpha}$, where $\alpha < 1.0$. This is also consistent with the notion that the stellar mass fraction increases from clusters to group-mass haloes \citep{gonzalez13,lagana13,budzynski14,vdB14}. In particular, \citet{viola15} measured the relation between halo mass and apparent richness for the GAMA groups. They found $M_{200}\propto N_{\mathrm{FoF}}^{1.09\pm0.18}$, which, after inverting $M$ and $N$, is consistent with the general picture, namely that lower-mass haloes  per unit halo mass host more bright satellites than higher-mass haloes do. We note, however, that this is measured using all FoF members associated with the group. This is quite different from the way we measured the UDG abundances. We proceed by making a direct comparison between the UDG abundance and the richness of GAMA groups in the same mass bins.

\subsection{Mass-richness relation of GAMA groups}\label{sec:richness}
To help place the measured abundances of UDGs into context, we measured the abundance of more massive galaxies (i.e.~richness) in the same groups. To make a direct comparison, we closely followed the analysis performed for the UDGs; we measured the richness within $R_{200}$, applied the same KiDS mask, and corrected the counts by assuming that the spatial distribution is uniform within $R_{200}$. To reduce the background correction compared to a purely photometric richness measurement, we made use of the full GAMA catalogue of all spectroscopically targeted galaxies. This is a valid approach given that the GAMA catalogue is essentially 100\% complete down to an $r$-band magnitude limit of 19.8, and does not suffer from sampling problems of close galaxy pairs thanks to multiple visits of each field. 

We measured richnesses down to a fixed magnitude past the characteristic magnitude in the $r$ band, which is $M^{*}_{r}=-21.21$ at $z=0.10$ \citep{blanton03}. The spectroscopic completeness limit of 19.8 in the $r$ band allowed us to measure the richness down to $M^{*}_{r} +2.5$ at this redshift. We k-corrected this limit based on a passively evolving stellar population with an age of 10 Gyr, and measured the richness consistently down to the same limit at all redshifts. 

\begin{figure}
\resizebox{\hsize}{!}{\includegraphics{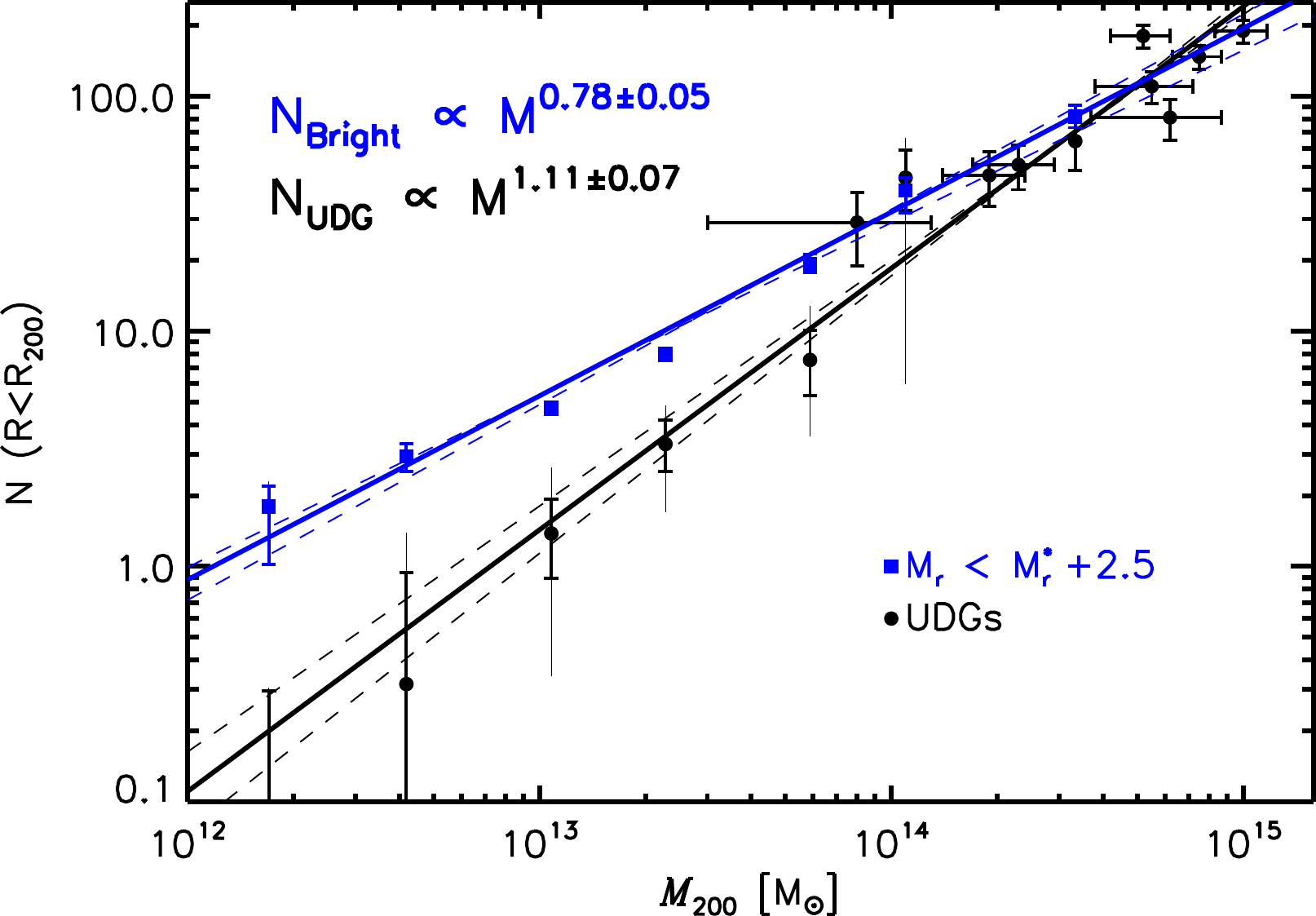}}
\caption{Abundance of UDGs as a function of halo mass. \textit{Thick black points with errors:} same as shown in Fig.~\ref{fig:abundance1}, and these include the cluster points from vdB16. \textit{Black lines:} best-fitting power-law relation including all black data points shown here. \textit{Blue squares:} richnesses (i.e.~the abundance of massive galaxies) in the same GAMA groups, with error bars defined in a similar way. \textit{Blue lines:} best-fitting power-law relation to the richness points.}
\label{fig:abundance3}
\end{figure}

We estimated the velocity dispersion $\sigma_{\mathrm{LOS}}$ of each group based on its $M_{200}$, assuming the scaling of \citet{evrard08}. Then we selected galaxies within 5$\sigma_{\mathrm{LOS}}$ from the group redshift, which is defined as the median redshift of all FoF members. We subtracted the background (i.e.~galaxies that are interlopers due to their redshift) statistically, measuring the density of galaxies in the same redshift slice (corresponding to $\pm5\sigma_{\mathrm{LOS}}$) in the same GAMA field (G09, G12, G15), but outside the group $R_{200}$. 

The data points are presented in Table~\ref{tab:abundance} and are shown in Fig.~\ref{fig:abundance3}. We fitted a power-law relation to these binned data points in the same way as was done for the UDG abundance measurements, finding a best-fit relation of 
\begin{equation}
N_{\mathrm{Bright}}=(31\pm3) \cdot \left[ \frac{M_{200}}{10^{14}\,\mathrm{M_{\odot}}}\right]^{0.78\pm0.05},
\end{equation}  
with $\chi ^2/\mathrm{d.o.f.}=4.2/5$.

The 5$\sigma$ cut is large enough to include all galaxies that are physically associated with the group, while it is small enough to not include substantial numbers of interlopers. A cut of $\pm3\sigma_{\mathrm{LOS}}$ is too strict to allow for all associated galaxies, especially for groups with low halo masses; in this case, we find the same best- fitting normalisation of $31\pm4$, but a slightly steeper slope of $0.86\pm0.06$. At the other extreme, we tested a fixed redshift cut of $\Delta z=$0.01. This conservative selection of group galaxies gives the same best-fit normalisation as the $\pm5\sigma$ cut and a similar slope of $0.79\pm0.05$. This shows that the richness measurement is robust with respect to the exact velocity cut along the line of sight. The slope is fully consistent with almost all relations measured by the literature studies that we listed above. 

The difference in trends between UDGs and more massive satellites (mass-richness relation), indicates that UDGs are not merely a fixed fraction of the total galaxy population that is independent of the environment. It is striking that in cluster-sized haloes the number of UDGs is approximately equal to the number of galaxies down to $M_{r} < M_{r}^{*} +2.5$. Conversely, for groups of $M_{200}\simeq 10^{12}\,\mathrm{M_{\odot}}$, the bright galaxies are a factor $\sim$10 more abundant than UDGs. Although the number of UDGs is comparable to these brighter galaxies in massive groups and clusters, we note that they still constitute only a small fraction of the total $r$-band luminosity (or stellar mass) even for cluster-sized systems. The total stellar mass contained in UDGs is about 1.0\% of the total stellar mass in galaxies with $M_{r} < M_{r}^{*} +2.5$ for clusters. This fraction drops further when compared to the total stellar mass (due to the contribution from fainter galaxies, and the intra-group and intra-cluster light), and is lower for groups.

Ultra-diffuse galaxies therefore seem to either form, or survive, more easily in the most massive haloes. However, they may still simply be a fixed fraction of the general population of dwarf galaxies that is independent of the environment. There are claims that the luminosity function of cluster galaxies has an upturn at low luminosities compared to the field \citep{popesso05}, although recent measurements do not reproduce this upturn \citep{depropris13,agulli14}. If the overall dwarf galaxy population is relatively more abundant in cluster-sized haloes, this could mean that UDGs simply follow the same trend. An interesting approach would be to measure the total dwarf abundance in the same GAMA groups for a direct comparison, but the background correction for small and faint galaxies prevents this with the current data set (see the discussion in vdB16).

Galaxies that are currently in massive galaxy clusters have on average spent their life in more overdense environments than galaxies that are currently in lower-mass systems. Whether a satellite galaxy can survive in a massive halo depends on their subhalo mass, concentration, orbit, and mass distribution of the host halo. While tidal forces are weaker in group-sized haloes than in clusters, mergers with other satellite galaxies are typically more common because relative velocities are lower in groups. It is hard to predict which physical process (tidal stripping or mergers with other galaxies) would be dominant in destroying UDGs in massive haloes, but it is plausible that elliptic orbits in clusters that avoid their centres are the orbits that make the UDGs survive longest. This may explain why UDGs are relatively more abundant in clusters than in groups.

A detailed comparison of the morphological properties of UDGs in different environments may also provide clues as to the possible formation mechanisms for UDGs. The median S\'ersic index, an often-used morphological indicator of the UDG candidates in the clusters studied by vdB16, is $n_\mathrm{S\acute{e}rsic}=1.4$ (this median value is unchanged when we only consider UDG candidates brighter than the surface brightness limits reached in the present study). Interestingly, after accounting for background UDG candidates, we find that the median S\'ersic index of UDGs in the GAMA groups is $n_\mathrm{S\acute{e}rsic}=2.2$, and only about 33\% have a S\'ersic index consistent with 1 (within uncertainties). This means that if we had selected UDGs in groups with a similar morphology as the UDGs found in clusters, the abundance would rise even more steeply towards higher-mass systems. This suggests that the formation mechanisms for UDGs, or their subsequent evolution, depends on their environment.

\section{Summary and outlook}\label{sec:conclusion}
We have expanded the work of vdB16 and measured the abundance of UDGs in group-sized haloes in the KiDS and GAMA surveys. The KiDS images are only several tenths of a magnitude shallower than the CFHT imaging stacks used in vdB16, which ensured a consistent comparison (after applying the corrections we described). The main conclusions of this study can be summarised as follows.

\begin{itemize}
\item The size distribution at a given surface brightness is steep, with a best-fitting power law of $n\, \mathrm{[dex^{-1}]}\propto r_\mathrm{{eff}}^{-2.7\pm 0.3}$. This is slightly shallower, but still roughly consistent (within $\sim 2\sigma$) to what has been found for UDGs in clusters by vdB16.
\item We were able to constrain the abundance of UDGs down to halo masses of $M_{200}\sim 10^{12}\,\mathrm{M_{\odot}}$. In such low-mass haloes, UDGs are rare; only about one  in ten of these groups contains a UDG that we can detect. The relation between the number of UDGs and the host halo masses, which now spans three orders of magnitude in halo mass and is well constrained,
is $N_{\mathrm{UDGs}} (\mathrm{R} < R_{200})\propto M_{200}^{1.11\pm0.07}$. Based on this relation, we expect about one in five systems like the Local Group to contain a UDG that we can detect. 
\item We also measured the mass-richness relation of more massive galaxies in a similar way, finding a significantly shallower relation: $N_{\mathrm{Bright}} (\mathrm{R} < R_{200})\propto M_{200}^{0.78\pm0.05}$. Interestingly, in cluster-sized haloes, the number of UDGs is approximately equal to the number of galaxies down to $M_{r} < M_{r}^{*} +2.5$. In contrast, for groups of $M_{200}\simeq 10^{12}\,\mathrm{M_{\odot}}$, the bright galaxies are a factor $\sim$10 more abundant than UDGs.
\item We find a median S\'ersic index of $n_\mathrm{S\acute{e}rsic}=2.2$ for UDGs in groups, versus $n_\mathrm{S\acute{e}rsic}=1.4$ for UDGs in clusters (vdB16). This is an interesting constraint on theoretical models that aim to explain the formation and survival of UDGs, as it may hint at different formation mechanisms, or a different subsequent evolution, in different environments.
\end{itemize}

The different relations between the UDG abundance and the richness of more massive galaxies versus halo mass suggest that UDGs are either destroyed more efficiently in group-sized haloes than in cluster-sized haloes, or that galaxy clusters may play a positive role in their creation. It may also be that the overall dwarf galaxy population is more abundant in clusters than in lower-mass systems, and that the UDGs constitute a fixed fraction of this general dwarf population. 
More detailed modelling is required to better interpret the differences measured here. 

On the observational side, progress can be made by performing detailed comparisons between the properties of UDGs in different environments (beyond comparing a median S\'ersic index). 
Unfortunately, the photometric data set we used is not ideal to investigate this; except for the $r$ band, which we used in this analysis, the imaging in the other filters is too shallow to measure meaningful colours. This prevents us from increasing the contrast of UDGs against the background, and we reach an overdensity of only 29\%. The Hyper Suprime-Cam survey, which recently released imaging over an overlapping sky region \citep{HSC17}, may be the way forward to study UDGs with deeper imaging in more detail. 

\begin{acknowledgements}
We thank Annie Robin, Emanuele Daddi, and Ryan Johnson for insightful discussions, and an anonymous referee for a constructive report with valuable points to improve and clarify this manuscript. The research leading to these results has received funding from the European Research Council under the European Union's Seventh Framework Programme (FP7/2007-2013) / ERC grant agreement n$^{\circ}$ 340519. H.Hoekstra acknowledges support from the European Research Council under FP7 grant number 279396. C.H. acknowledges support from the European Research Council under grant number 647112. H.Hildebrandt is supported by an Emmy Noether grant (No. Hi 1495/2-1) of the Deutsche Forschungsgemeinschaft. D.K. is supported by the Deutsche Forschungsgemeinschaft in the framework of the TR33 `The Dark Universe'. K.K. acknowledges support by the Alexander von Humboldt Foundation. R.N. acknowledges support from the German Federal Ministry for Economic Affairs and Energy (BMWi) provided via DLR under project no. 50QE1103.

Based on data products from observations made with ESO Telescopes at the La Silla Paranal Observatory under programme IDs 177.A-3016, 177.A-3017 and 177.A-3018, and on data products produced by Target/OmegaCEN, INAF-OACN, INAF-OAPD and the KiDS production team, on behalf of the KiDS consortium. OmegaCEN and the KiDS production team acknowledge support by NOVA and NWO-M grants. Members of INAF-OAPD and INAF-OACN also acknowledge the support from the Department of Physics \& Astronomy of the University of Padova, and of the Department of Physics of Univ. Federico II (Naples).

GAMA is a joint European-Australasian project based around a spectroscopic campaign using the Anglo-Australian Telescope. The GAMA input catalogue is based on data taken from the Sloan Digital Sky Survey and the UKIRT Infrared Deep Sky Survey. Complementary imaging of the GAMA regions is being obtained by a number of independent survey programmes including GALEX MIS, VST KiDS, VISTA VIKING, WISE, Herschel-ATLAS, GMRT and ASKAP providing UV to radio coverage. GAMA is funded by the STFC (UK), the ARC (Australia), the AAO, and the participating institutions. The GAMA website is http://www.gama-survey.org/.
\end{acknowledgements}

\bibliographystyle{aa} 
\bibliography{MasterRefs} 

\begin{appendix}
\section{Robustness tests}\label{sec:robustness}\label{sec:robustnesssize}
\begin{figure}
\resizebox{\hsize}{!}{\includegraphics{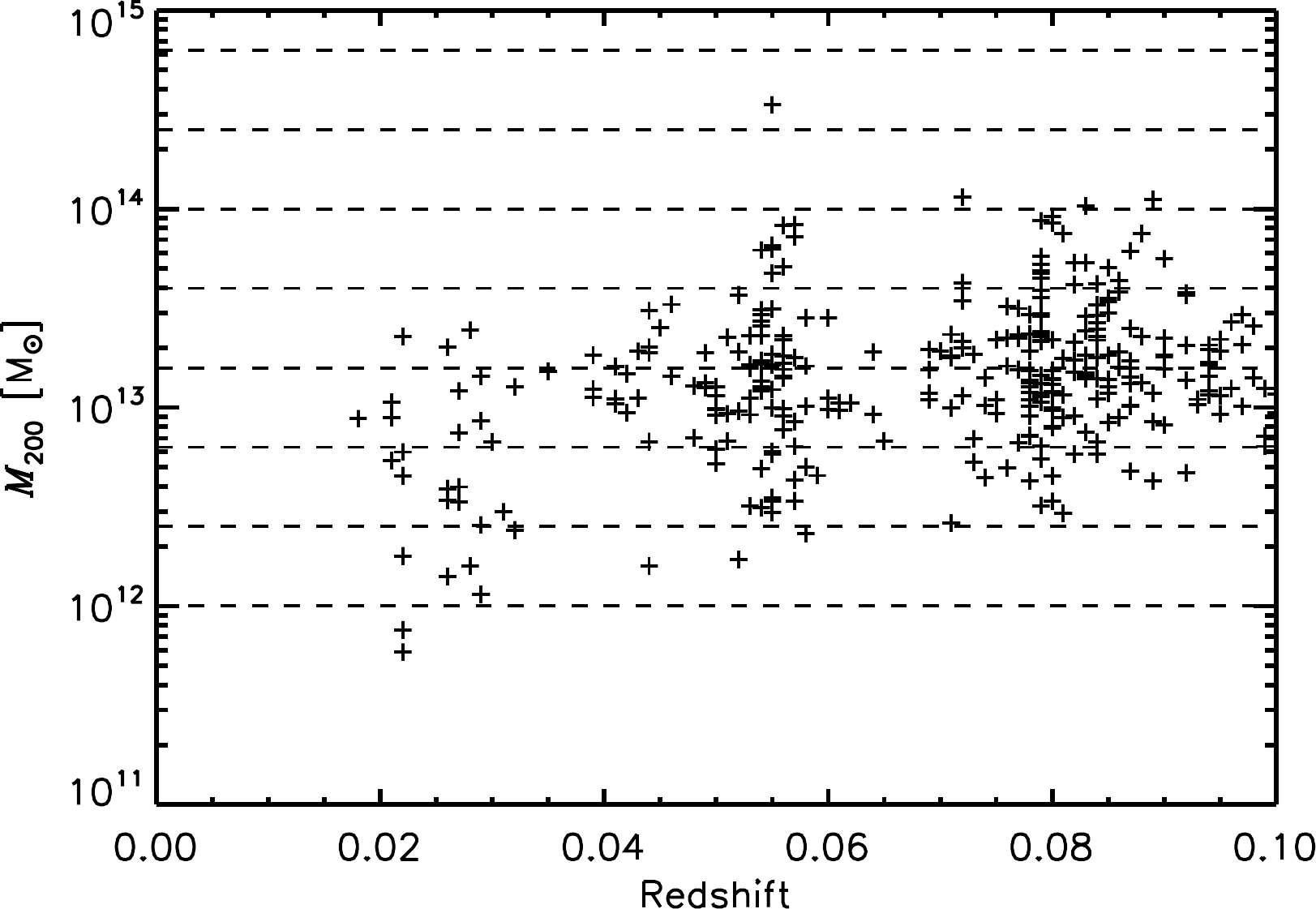}}
\caption{Distribution in redshift and mass of the 325 GAMA galaxy groups we studied. The mass bins we used for the abundance measurements are indicated.} 
\label{fig:redshiftmass}
\end{figure}

Here we discuss the interdependence of the two main measurements in this paper, namely the abundance of UDGs as a function of halo mass and their size distribution, and assess the robustness of each individual measurement. We would expect a correlation between redshifts and halo masses of the galaxy groups because
the probed volume is larger at high-$z$ and because of the higher sensitivity at low-$z$ (since the limiting magnitude of the spectroscopic target selection probes further down the luminosity function at low-$z$). This correlation is the reason for this interdependence. For instance, the correction factors described in Sect.~\ref{sec:analysis_correctionfactors} are mainly a function of redshift, but depend on the assumed size distribution. Since Fig.~\ref{fig:abundance1} shows the main results in bins of mass, any correlation between halo masses and redshifts of groups would create an uncertainty in the correction factors (or an incorrectly assumed size distribution) appear as a tilt in the abundance versus mass relation. The good news is that although a correlation between redshift and halo mass exists for our sample, it is moderate in this redshift range 
\citep[Fig.~\ref{fig:redshiftmass}, and for the distribution over a larger redshift baseline, cf. Fig.~16 in][]{robotham11}. 
This means that the results do not sensitively depend on each other, as we illustrate with several tests below.

Our fiducial measurement of the abundance-mass relation assumes a UDG size power-law with exponent=-2.70. This leads to a measured abundance power-law=$1.11 \pm 0.07$ (the main result presented in Sect.~\ref{sec:abundancehalomass}). When we change the assumed UDG size power-law to exponent=-3.0 or -2.4 (i.e. 1$\sigma$ lower or higher than the fiducial model), we measure an abundance power-law exponent of $1.10 \pm 0.07$ and $1.13 \pm 0.07$, respectively. Even when we take the steeper size measured in vdB16 (size power-law exponent of -3.40), we measure a moderate change in the abundance power-law exponent ($1.08 \pm 0.07$). This shows that the assumed UDG size distribution does not significantly affect the measured UDG abundance versus halo mass. 

As described in the main text, we have assumed a weak-lensing calibration of the total mass associated with the galaxy groups, based on their total $r$-band luminosity, as outlined in \citet{viola15}. We explore what would happen if we were to very conservatively increase or decrease the masses of the galaxy groups by 20\%. This results in a measured abundance power-law exponent of $1.10 \pm 0.06$ and $1.07 \pm 0.06$, respectively. 
We can conclude that uncertainties on the average GAMA group mass measurements do not affect our results.

The results presented in this work rely on a statistical background correction, which is substantial for the marginally overdense groups studied here. There are different ways of performing this background correction, and as explained in the text, we used the average density of UDG candidates (with angular sizes $1\farcs5\leq r_\mathrm{eff}\leq 8\farcs0$) over the entire KiDS region. If we were to measure the density outside of the projected virial radii of all GAMA groups up to redshift $z\leq 0.10$, we would measure an abundance power-law with exponent $1.09 \pm 0.07$. Our results are thus robust with respect to the details of the background subtraction method.

\begin{figure}
\resizebox{\hsize}{!}{\includegraphics{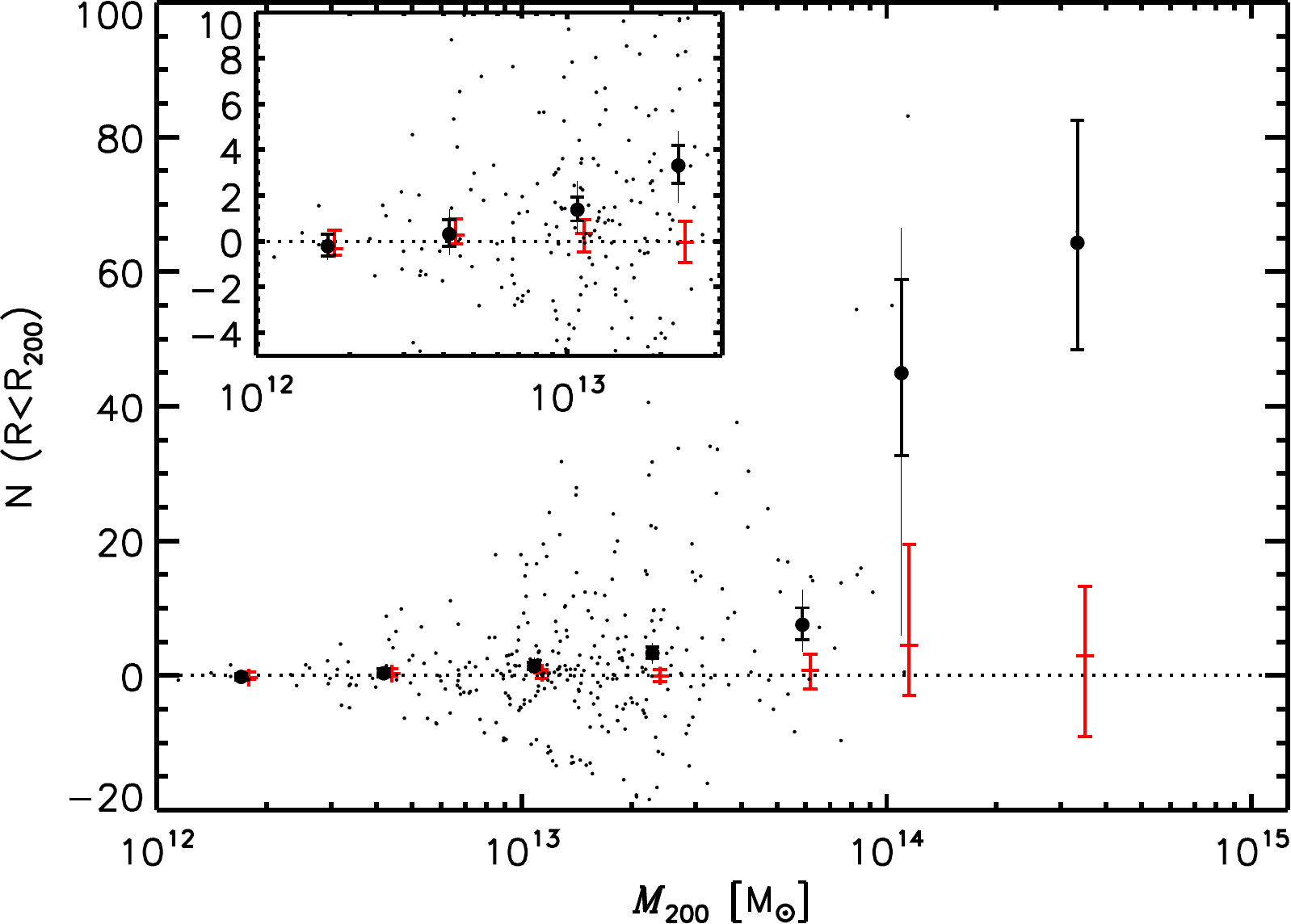}}
\caption{Abundance of UDGs as a function of halo mass. \textit{Small black dots:} individual measurements per group (using real data). \textit{Thick black points and errors:} identical to that shown in Fig.~\ref{fig:abundance1}. \textit{Red points:} mean UDG number densities per group when the UDGs are injected at random positions into the KiDS images, with 68\% confidence interval from bootstrapping random samples. The inset shows a zoomed-in region. The abundance measurements show a clear signal, while the ``randoms'' are consistent with zero.}
\label{fig:abundance_nul}
\end{figure}

Another sanity check we performed is to repeat the same analysis using the three million simulated sources that were injected at random positions into the KiDS images. We made 25 random resamplings of the simulated catalogue, where we gave a probability for each source to be picked based on their sizes. In this way, every resampled catalogue has the same size distribution as the observed size distribution of UDG candidates. Each realisation thus contains 16\,595 sources ($\pm$ Poisson noise). The median measurements, with 68\% confidence regions, are shown by the red points in Fig.~\ref{fig:abundance_nul}. This shows that the signal is consistent with zero, with a spread between the 25 realisations that is roughly equal to the estimated uncertainties on the real data points. 

If bright group galaxies, or intra-group light (IGL), had played a significant role in defining the sample of UDGs (in particular, \textit{\textup{preventing}} us from detecting UDGs close to the group centres), this would have shown up as a deficit in the random samples, which is not the case here. 

We note that there are three groups in the mass bin $10^{14.0}\leq M_{200}/\mathrm{M_{\odot}}\leq 10^{14.4}$, but one of them has a very marginal overdensity compared to the background. This results in a very large bootstrap error.

The measured relation between the abundance and mass of groups in turn affects the estimated size distribution. First we recall that the smallest UDGs that are plotted in Fig.~\ref{fig:sizedist} are not directly measured in the highest-$z$ groups because their angular sizes are then smaller than $1\farcs5$. The fitted power-law size distribution thus already enters into the correction factor. Moreover, we normalised the distribution here per ``virial disk'' ($\pi \cdot R_{200}^2$). Since the number of UDGs is a strong function of the halo mass (cf.~\ref{fig:abundance1}), primarily probing high-mass haloes at high redshift would introduce an artificial tilt in the measured relation. To circumvent this, we scaled the UDG numbers by dividing them over $M_{200}^{1.1}$. Again, if we were to probe the same halo mass distribution at different redshifts, such a scaling would have no effect. 

We tested the effect of our assumed abundance versus halo mass relation on the estimated size distribution. Our fiducial assumption was a scaling by $M_{200}^{1.1}$, which led to a size power-law exponent of $-2.71\pm 0.33 $. If we were to assume an abundance scaling of $M_{200}^{1.3}$ or $M_{200}^{0.9}$ (i.e. 3$\sigma$ higher or lower than the fiducial model), we find $-2.70 \pm 0.34$ and $-2.70 \pm 0.33,$ respectively. Uncertainties in the assumed abundance-mass distribution thus have a negligible effect on the measured size distribution.

\section{Examples}\label{sec:examples}
\begin{figure*}
\resizebox{\hsize}{!}{\includegraphics{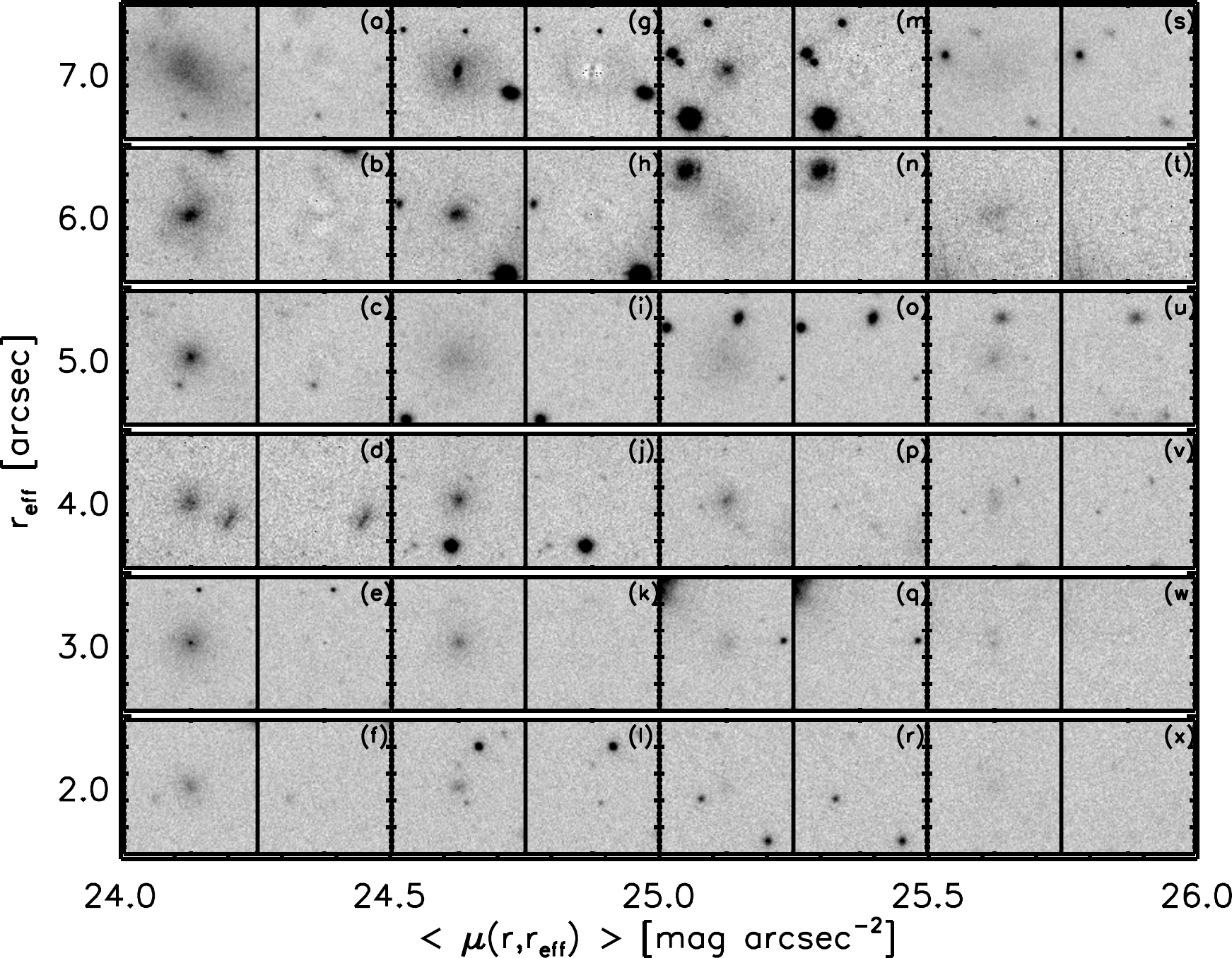}}
\caption{Example of typical galaxies we selected in different parts of parameter space (angular size versus mean surface brightness within the effective radius). \textit{Left panels:} original $r$-band images, spanning 21$''\times$21$''$ centred on the galaxy. \textit{Right panels:} residual of the $r$-band image after model subtraction. The best-fit morphological parameters are listed in Table~\ref{tab:exampleparameters}.}
\label{fig:cutouts_paper}
\end{figure*}

To provide a qualitative sense of the depth of the KiDS imaging, Fig.~\ref{fig:cutouts_paper} shows selected UDG candidates in different parts of parameter space (as described in Sect.~\ref{sec:selectioncriteria} and shown in Fig.~\ref{fig:completeness}). The best-fitting S\'ersic parameters are listed in Table~\ref{tab:exampleparameters}. Completeness limits are quantified in Sect.~\ref{sec:simulations}.

\begin{table*}
\caption{Best-fitting \texttt{GALFIT} parameters of the examples shown in Fig.~\ref{fig:cutouts_paper}.}
\label{tab:exampleparameters}
\begin{center}
\begin{tabular}{l c c c | l c c c | l c c c }
\hline
\hline
&$\mathrm{r_{eff}}$&$\langle\mu(r,\mathrm{r_{eff}})\rangle$ & S\'ersic &&$\mathrm{r_{eff}}$&$\langle\mu(r,\mathrm{r_{eff}})\rangle$ & S\'ersic &&$\mathrm{r_{eff}}$&$\langle\mu(r,\mathrm{r_{eff}})\rangle$ & S\'ersic \\
& [arcsec]  &  [mag arcsec$^{-2}$]  &n && [arcsec]  &  [mag arcsec$^{-2}$]  &n && [arcsec]  &  [mag arcsec$^{-2}$]  &n \\
\hline
(a)&7.31&24.10&0.96&(i)&5.41&24.93&1.04&(q)&2.74&25.42&1.95\\
(b)&5.75&24.26&2.11&(j)&4.12&24.69&2.86&(r)&1.61&25.39&3.09\\
(c)&4.67&24.49&2.63&(k)&2.62&24.53&1.94&(s)&6.77&25.67&0.87\\
(d)&3.62&24.44&2.00&(l)&1.73&24.50&2.29&(t)&6.00&25.56&1.83\\
(e)&3.35&24.23&1.50&(m)&6.86&25.26&3.98&(u)&4.88&25.78&2.53\\
(f)&1.75&24.13&1.10&(n)&5.54&25.01&0.75&(v)&3.70&25.73&2.34\\
(g)&6.72&24.58&3.43&(o)&4.71&25.06&0.80&(w)&2.72&25.76&2.71\\
(h)&6.50&24.97&3.45&(p)&3.72&25.00&2.74&(x)&1.95&25.70&1.14\\
\hline
\end{tabular}
\end{center}
\end{table*}

\end{appendix}

\end{document}